\documentclass[journal]{IEEEtran}   
\usepackage{multirow}
\usepackage{diagbox}
\usepackage{amssymb}
\usepackage{amsmath}
\usepackage{graphicx}
\usepackage{cite}
\usepackage{subfigure}
\usepackage{graphicx,epstopdf}
\usepackage{epsfig}	
\usepackage{cite,graphicx,amsmath,amssymb}
\usepackage{comment}

\usepackage{amssymb}
\usepackage{amsmath}
\usepackage{url}
\usepackage{xcolor}
\usepackage{cite,graphicx,amsmath,amssymb}
\usepackage{subfigure}
\usepackage{fancyhdr}
\usepackage{mdwmath}
\usepackage{mdwtab}
\usepackage{caption}
\usepackage{amsthm}
\usepackage{lipsum}
\usepackage{float}
\usepackage{graphicx}
\usepackage{flafter}
\usepackage[section]{placeins}
\usepackage{tabularx}
\usepackage{array}
\usepackage[numbers,sort&compress]{natbib}
\usepackage{makecell}
\setcitestyle{square,citesep={,}}

\newtheorem{theorem}{Theorem}

\newtheorem{lemma}{Lemma}

\newtheorem{corollary}{Corollary}

\newtheorem{remark}{Remark}  

\makeatletter
\def\ScaleIfNeeded{%
\ifdim\Gin@nat@width>\linewidth \linewidth \else \Gin@nat@width
\fi } \makeatother

\usepackage[hidelinks]{hyperref}

\newcommand{\Remarkref}[1]{%
  \hyperref[#1]{\textbf{Remark~\ref*{#1}}}}

\newcommand{\Corollaryref}[1]{%
  \hyperref[#1]{\textbf{Corollary~\ref*{#1}}}}

\newcommand{\Figref}[1]{%
  \hyperref[#1]{Fig.~\ref*{#1}}}

\begin{document}

\title{\Huge{Exploiting Pinching Antenna Systems in NOMA Covert Communication Networks}}

\author{Xinwei~Yue,~\IEEEmembership{Senior Member,~IEEE}, Hao Li,~\IEEEmembership{Graduate Student Member,~IEEE}, Jingjing Zhao,~\IEEEmembership{Senior Member,~IEEE}, Xianfu~Lei,~\IEEEmembership{Member,~IEEE}, Yuanwei\ Liu,~\IEEEmembership{Fellow, IEEE}
\thanks{X. Yue and H. Li are with the College of Information and Communication Engineering, and with the Center for Target Cognition Information Processing Science and Technology, Beijing Information Science $\&$ Technology University, and also with the Key Laboratory of Modern Measurement $\&$ Control Technology, Ministry of Education, Beijing Information Science $\&$ Technology University, Beijing 102206, China (email: \{xinwei.yue and lihao\}@bistu.edu.cn).}%
\thanks{Y. Liu is with the Department of Electrical and Electronic Engineering, The University of Hong Kong, Pokfulam, Hong Kong (email: yuanwei@hku.hk).}%
\thanks{J. Zhao is with the School of Electronic and Information Engineering,
Beihang University, Beijing 100191, China
(e-mail: jingjingzhao@buaa.edu.cn).}
\thanks{X. Lei is with the School of Information Science and Technology, Southwest Jiaotong University, Chengdu 611756, China (e-mail: xflei@swjtu.edu.cn).}
}

\maketitle

\begin{abstract}
Pinching antenna systems (PASS) provide flexible transmitter-side radiation reconfiguration for high-frequency wireless communications. This paper applies PASS to non-orthogonal multiple access (NOMA) covert communication networks, where activated pinching antennas (PAs) transmit superimposed signals to a pair of users stochastically distributed within a rectangle. Two passive wardens respectively monitor the downlink transmissions intended for two users. More specifically, we derive the closed-form detection error probability (DEP) expressions for two wardens under bounded noise uncertainty. Then we analytically determine the optimal detection thresholds and minimum DEPs in the worst-case covert detection scenarios. The covert transmission rates of two users are further characterized by taking into consideration ideal successive interference cancellation and non-ideal successive interference cancellation at near user and intra-NOMA interference at far user. Numerical results are presented to confirm that: 1) PASS-NOMA is capable of providing higher covert rates relative to conventional orthogonal multiple access OMA schemes; 2) As the number of activated PAs begun to increase, the near user has ability to obtain a larger covert rate gain than far user; and 3) The DEP of PASS-NOMA networks is larger than that of both PASS-OMA and traditional antenna based NOMA in specific output power regions.
\end{abstract}


\section{Introduction}

Sixth-generation (6G) wireless networks are expected to support wide-area coverage, ultra-high data rates, and ubiquitous connectivity \cite{Saad2020Network}. Although the high-frequency bands provide abundant bandwidth for high-rate transmission, the resulting links are vulnerable to severe propagation loss, blockage, and line-of-sight (LoS) sensitivity to user locations and link geometry \cite{Giordani2020MCOM,Tataria2021PIEEE}. To address these propagation challenges, pinching-antenna systems (PASS) have emerged as a flexible antenna architecture for high-frequency wireless transmission \cite{Liu2025MCOM}. Specifically, the basic PASS architecture consists of a dielectric waveguide equipped with specialized dielectric elements, which enable guided electromagnetic waves to radiate into free space \cite{Fukuda2022DOCOMO}. These elements create controllable radiation points along the waveguide, each functioning as a pinching antenna (PA). By activating one or more PAs, PASS is able to provide spatial adaptability at the transmitter side and reshapes the corresponding LoS channels \cite{Ding2025TCOM}.

Recent studies on PASS have progressed from fundamental array analysis to multi-user and multi-waveguide transmission \cite{Ouyang2025PASSArrayGain,Wang2025PASSBeamforming}. The gain of PASS was characterized in \cite{Ouyang2025PASSArrayGain}, which reveals how the number and spacing of PAs affect the attainable gain. To minimize transmit power, the authors of \cite{Wang2025PASSBeamforming} developed a physics-based PASS model and jointly designed transmit and pinching beamforming. For uplink transmission, the ergodic rate expressions of PASS were derived to characterize the performance by fixing the position of PAs \cite{Hou2026PASSUplink}. In multi-user uplink PASS scenarios \cite{Bereyhi2026MIMOPASS}, hybrid multi-user detection was investigated to improve the uplink transmission performance. The segmented waveguide structure with optimized PA placement was proposed to improve the uplink signal-to-noise ratio \cite{Ouyang2026SWAN}. For downlink transmission, the location of PAs were optimized to maximize the achievable rate \cite{Xu2025PASSRateMax}. Extending this design to downlink PASS with multiple users, transmit and pinching beamforming were jointly optimized for the purpose of sum-rate maximization \cite{Xu2026PASSJointBeamforming}. Under imperfect channel state information (CSI), robust baseband and pinching beamforming were jointly designed to maximize the worst-case rate of downlink PASS \cite{Sun2026PASSRobust}. Beyond rate-oriented optimization, the spectral–energy efficiency tradeoff was investigated for single-user and multiuser PASS subject to quality-of-service constraints \cite{Zhou2026PASSSEEE}. In addition, the application of PASS to multicast communications was surveyed to maximize the achievable multicast rate \cite{Shan2026PASSMulticast}. The practical multi-waveguide transmission structures of PASS were proposed to solve the max-min fairness problems of multicast communications \cite{Zhao2026PASSMultiUser}. 

While existing researches have shown that PASS holds great potential for improving the security of transmission and enhancing link reliability \cite{Badarneh2026PASSPLS,Shan2026PASSSecureMulticast,Wang2026PASSPLS}. In \cite{Badarneh2026PASSPLS}, the average secrecy capacity and outage probability of PASS were analyzed by invoking single activated PA. With the emphasis on multicast transmission, the transmit and pinching beamforming of PASS were jointly optimized to maximize the secrecy multicast rate \cite{Shan2026PASSSecureMulticast}. The authors of \cite{Wang2026PASSPLS} designed the Sapley-value based PASS to improve the secrecy rate by taking into account phase adjustments. Moreover, the artificial noise aided pinching beamforming of PASS was  highlighted to improve the secrecy rates \cite{Zhu2026PASSSecure}. Considering imperfect CSI, the authors in \cite{Zhao2026PASSSecureRobust} formulated the maximization problems of sum rates by jointly artificial noise, PA power allocation and positions. The index modulation based PASS was elucidated to achieve secure transmission \cite{Zhong2026PASSIDM}. On the other hand, covert communication represents another secure transmission paradigm that aims to conceal the existence of wireless transmission from a warden. In \cite{Zhang2025PASSCovert}, the PA positioning and beamforming of PASS were jointly optimized to improve the covert transmission rate. For more complex propagation scenarios, covert communication was extended to near-field multi-waveguide PASS \cite{Agullana2026PASSNearFieldCovert}, where both PA placement and beamforming were jointly designed. Furthermore, sensing-aided PASS was investigated for mobile-warden scenarios by tracking the warden's CSI for covert transmission \cite{Jiang2025PASSCovertSensing}.

As a type of flexible antenna, reconfigurable intelligent surface (RIS) has been introduced to solve non-orthogonal multiple access (NOMA) secure communication problems \cite{Zhang2022IRSNOMASecurity}. On the condition of imperfect CSI, the robust secure transmission for RIS-NOMA networks was further investigated with hybrid beamforming \cite{Zhang2024IRSNOMARobust}. As a further advance, the secrecy performance of active RIS-NOMA networks was analyzed in terms of secrecy outage probability and system throughput \cite{Li2024ARISNOMA}. To achieve comprehensive security coverage, the authors of \cite{Xie2024STARNOMASecrecy} evaluated the average secrecy capacity of simultaneously transmitting and reflecting surface (STARS) aided NOMA networks. In addition to considering safety performance, the covert communication problems of RIS-NOMA networks were surveyed in \cite{Lv2022TWC} by exploiting RIS phase-shift uncertainty. The sensing and covert communication of RIS-NOMA was jointly discussed with uncertain warden locations \cite{Lei2026RISNOMACovertSensing}. The authors of \cite{Xiao2024TVT} studied the covert rate of STARS-NOMA networks and obtained the optimal detection threshold. Based on this, the covert transmission performance of active STARS aided-NOMA was analyzed by considering signal amplification and internal thermal noise \cite{Yue2025ASTARSCOVERT}.

The integration of PASS with NOMA networks represents a promising approach toward enhancing multi-user transmission flexibility in future wireless communications \cite{Wang2025WCL,Yue2026PASSNOMA,Xu2025TCOM}. In \cite{Wang2025WCL}, an antenna activation strategy was developed for PASS-NOMA by selecting PAs from predefined candidate positions to maximize the achievable sum rate. The authors of \cite{Yue2026PASSNOMA} further analyzed the outage probability and ergodic rate of PASS-NOMA under randomly distributed user locations with non-ideal successive interference cancellation (NISIC). To satisfy user rate requirements, PA positions and power allocation were jointly optimized for PASS-NOMA transmission \cite{Xu2025TCOM}. The multi-user PASS-NOMA framework was studied in \cite{Xue2026PASSNOMAResource}, which jointly optimized power allocation, pinching beamforming, and user scheduling to maximize the sum rate. Additionally, a joint transmit and pinching beamforming scheme was designed for multi-user PASS-NOMA systems to minimize the transmit power \cite{Gan2026PASSNOMA}. From the perspective of reliable transmission, the secrecy performance of PASS-NOMA with internal eavesdropping was analyzed by adjusting the PA coupling length \cite{Chi2025PASSNOMA}. 

\subsection{Motivations and Contributions}

As discussed above, existing studies on PASS-NOMA have mainly focused on transmission design, performance analysis, and resource optimization, whereas its application to covert communications remains relatively underexplored. Notably, PASS can flexibly adjust the spatial configuration of activated PAs to reshape the effective propagation characteristics, thereby providing additional degrees of freedom for covert transmission. Based on this feature, the work in \cite{Zhang2025PASSCovert} investigated PASS-assisted covert communications but did not consider NOMA transmission. In addition, the application of  PASS to covert communications by taking into consideration near-field multi-waveguide case \cite{Agullana2026PASSNearFieldCovert}, while the warden-side detection performance was not explicitly characterized. Since practical successive interference cancellation (SIC) imperfections introduce residual interference and thereby affect the covert transmission performance of PASS-NOMA networks, we explicitly consider both ideal SIC (ISIC) and NISIC. Motivated by these observations, we investigate a PASS-NOMA covert communication networks with two randomly distributed NOMA users and two passive wardens. The activated PAs transmit superimposed signals to users $r$ and $t$, while wardens $w_r$ and $w_t$ monitor the transmissions intended for the corresponding users, respectively. More specifically, we characterize the detection error probability (DEP) of each warden under bounded noise power uncertainty and further investigate the covert transmission rates of users $r$ and $t$ under the considered SIC conditions. The main contributions of this paper are summarized as follows:
﻿
\begin{enumerate}
\item We establish a two-warden analytical framework for PASS-NOMA covert communication networks, where wardens $w_r$ and $w_t$ monitor the transmissions intended for users $r$ and $t$, respectively. We further account for the asymmetric NOMA power coefficients and distinct PAs-to-warden leakage gains to characterize the detection performance of the two wardens separately, thereby revealing the coupling between PASS induced transmission gains and warden-side detection.
﻿
\item We investigate the false alarm (FA) and miss detection (MD) probabilities of PASS-NOMA and compare them with those of PASS based orthogonal multiple access (PASS-OMA) and traditional antenna based (TA-NOMA). We show that the FA probabilities of PASS-NOMA increase more rapidly with output power than those of TA-NOMA, whereas the FA probability of PASS-OMA remains nearly unchanged. We further demonstrate that the MD probabilities of PASS-NOMA decrease more rapidly with output power than those of the two benchmark schemes.
﻿
\item We derive the piecewise closed-form expressions for the DEP of both wardens and further obtain the optimal detection thresholds and corresponding minimum DEPs. We demonstrate that PASS-NOMA achieves higher DEP than PASS-OMA and TA-NOMA within specific output power ranges. We reveal that varying the number of activated PAs changes both the DEP level and the location of its minimum, while larger noise power uncertainty enlarges the output power range with higher DEP values.
﻿
\item We derive the approximated expressions for covert transmission rates of user $r$ with ISIC/NICIC and user $t$. We demonstrate that PASS-NOMA achieves much higher covert transmission rates than conventional single-antenna NOMA. We observe that increasing the number of activated PAs, PASS-NOMA has ability to provided the larger covert transmission rate for user with ISIC, whereas the rate of user $t$ approaches an interference-limited ceiling at high transmit power.

\end{enumerate}

\subsection{Organization and Notation}

The remainder of this paper is organized as follows. Section~II presents the system model of PASS-NOMA covert communication networks. Section~III derives the DEP expressions under bounded noise uncertainty. Section~IV characterizes the covert transmission rates under different SIC conditions. The numerical results and analyses are presented in Section V. Finally, Section~VI concludes the paper. The proofs of the main analytical results are provided in the appendices.

The main notations used in this paper are summarized as follows. For a random variable $X$, $f_X(\cdot)$ and $F_X(\cdot)$ denote its probability density function and cumulative distribution function, respectively. The operators $\mathbb{E}\{\cdot\}$ and $\Pr(\cdot)$ denote expectation and probability, respectively. The superscripts $(\cdot)^T$ and $(\cdot)^H$ represent the transpose and conjugate transpose, respectively. The notation $|\cdot|$ denotes the absolute value of a scalar, whereas $\|\cdot\|$ denotes the Euclidean norm of a vector or matrix.

\section{System Model}\label{System Model}
\begin{figure}[t!]
    \begin{center}
        \includegraphics[width=3.8in,  height=2.4in]{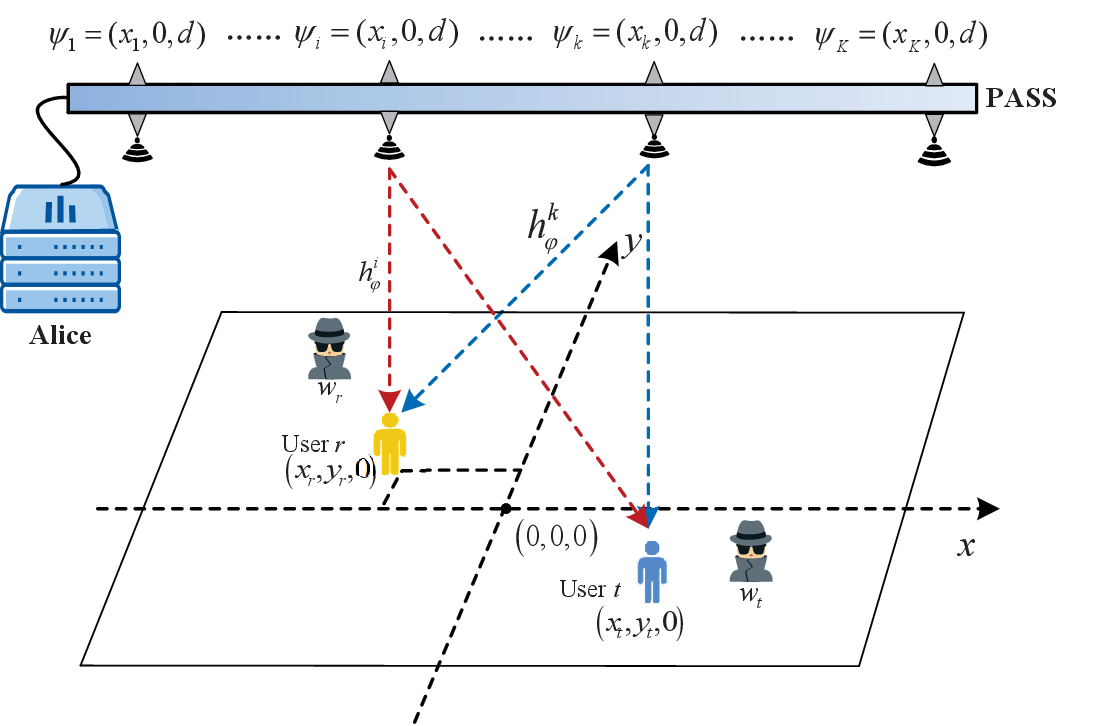}
        \caption{Illustration of exploiting PASS in NOMA covert communication networks.} 
        \label{Fig. 1}
    \end{center}
\end{figure}

We consider one PASS aided downlink NOMA covert communication scenaries, as shown in Fig.~1. where the BS equipped with PASS transmits the superimposed signals to a couple of single-antenna users, namely the near user $r$ and far user $t$. Specifically, the BS is located at the edge of a rectangular service region with coordinate $\mathbf{B}=(-D_1,0,0)$, and both the length and width of the service region are $2D_1$. The PASS is deployed parallel to the ground at a height of $d$, where $K$ PAs are mounted on one dielectric waveguide to assist downlink transmission \cite{Wang2025WCL}. In the presence of legitimate transmissions, two passive wardens, denoted by $w_r$ and $w_t$, attempt to monitor the transmission activities associated with users $r$ and $t$, respectively \cite{Lv2022TWC}. At this moment, $w_r$ aims to detect whether the message intended for user $r$ is transmitted, whereas $w_t$ aims to detect the transmission associated with user $t$. The users and wardens are assumed to be distributed on the ground plane. A common narrowband transmission model is adopted, and all wireless links are assumed to be dominated by LoS propagation, which are consistent with high-frequency PASS-assisted communication systems.

\subsection{Signal Model}

With help of PASS, the BS transmits the superposed signals to user $\varphi$, i.e., $\varphi\in\{r,t\}$, which can be expressed as
\begin{equation}
y_{\varphi}
=
\mathbf{h}_{\varphi}^{H}\mathbf{w}
\left(
\sqrt{a_r}x_r+\sqrt{a_t}x_t
\right)
+n_{\varphi},
\end{equation}
where $x_r$ and $x_t$ denote the normalized information symbols intended for users $r$ and $t$, respectively, satisfying $\mathbb{E}\{|x_r|^2\}=\mathbb{E}\{|x_t|^2\}=1$, while $a_r$ and $a_t$ denote the corresponding power allocation coefficients satisfying $a_r+a_t=1$ and $a_r<a_t$. $n_{\varphi}\sim\mathcal{CN}(0,\sigma_{\varphi}^{2})$ represents the additive white Gaussian noise at user $\varphi$. In addition, $\mathbf{h}_{\varphi}=
\left[
\frac{\sqrt{\eta}e^{-j\frac{2\pi}{\lambda}\|\boldsymbol{\psi}_{\varphi}-\boldsymbol{\psi}^{\rm PA}_1\|}}
{\|\boldsymbol{\psi}_{\varphi}-\boldsymbol{\psi}^{\rm PA}_1\|},
\ldots,
\frac{\sqrt{\eta}e^{-j\frac{2\pi}{\lambda}\|\boldsymbol{\psi}_{\varphi}-\boldsymbol{\psi}^{\rm PA}_K\|}}
{\|\boldsymbol{\psi}_{\varphi}-\boldsymbol{\psi}^{\rm PA}_K\|}
\right]^T$
denotes the channel vector from $K$ PAs to user $\varphi$, with $\boldsymbol{\psi}_{\varphi}$ and $\boldsymbol{\psi}^{\rm PA}_i$ denoting the locations of user $\varphi$ and the $i$-th PA, respectively. Here, $\eta=\lambda^2/(16\pi^2)$ and $\lambda$ represents the free-space carrier wavelength. Moreover, $\mathbf{w}=
\left[
\sqrt{p_1}e^{-j\theta_1},
\ldots,
\sqrt{p_K}e^{-j\theta_K}
\right]^T$ denotes the transmission weight vector accounting for the power allocation and phase progression along the waveguide, where $p_i=P_b/K$ is the transmit power allocated to the $i$-th PA, $P_b$ denotes the total transmit power of the BS and $
\theta_i=\frac{2\pi\|\boldsymbol{\psi}^{\rm PA}_i-\boldsymbol{\psi}^{\rm PA}_0\|}{\lambda_g}$
is the phase shift introduced by signal propagation inside the waveguide from the common feed point $\boldsymbol{\psi}^{\rm PA}_0$ to the $i$-th PA.
Furthermore, the guided wavelength $\lambda_g$ inside the dielectric waveguide is given by $\lambda_g=\lambda/n_{\rm eff}$, with $n_{\rm eff}$ denoting the effective refractive index \cite{Zhang2025PASSCovert}.

Following the NOMA decoding order, user $r$ first decodes the message intended for user $t$ and then applies SIC to recover its own signal. The corresponding SINR at user $r$ for decoding the message of user $t$ is given by
\begin{equation}
\gamma_{r\rightarrow t}
=
\frac{
a_t |\mathbf{h}_r^H\mathbf{w}|^2
}{
a_r |\mathbf{h}_r^H\mathbf{w}|^2+\sigma_r^2
}.
\end{equation}

After SIC, the SINR at user $r$ for decoding its own signal can be expressed as
\begin{equation}
\gamma_r
=
\frac{
a_r |\mathbf{h}_r^H\mathbf{w}|^2
}{
\vartheta \rho |h_I|^2+\sigma_r^2
},
\end{equation}
where $h_I$ represents the residual interference channel coefficient under NISIC and $\rho$ represents the residual interference power coefficient. $\vartheta\in\{0,1\}$ is a binary indicator of the SIC condition that $\vartheta=0$ and $\vartheta=1$ corresponding to ISIC and NISIC, respectively. User $t$ directly detects its own information symbol $x_t$ by regarding the signal intended for user $r$ as intra-NOMA interference. The corresponding SINR is given by
\begin{equation}
\gamma_t
=
\frac{
a_t |\mathbf{h}_t^H\mathbf{w}|^2
}{
a_r |\mathbf{h}_t^H\mathbf{w}|^2+\sigma_t^2
}.
\end{equation}

\section{Detection Error Probability}

In this section, we investigate the DEP at the wardens. Owing to the passive and non-cooperative nature of the wardens, the legitimate transceivers cannot accurately know the exact noise power at the wardens and only characterize it through a bounded uncertainty interval. To capture this practical effect, the bounded noise uncertainty model is adopted, where the noise power $
\sigma_{w_\varphi,\mathrm{dB}}^2$ at warden $w_\varphi$ is assumed to lie within a bounded interval, i.e., $
\sigma_{w_\varphi,\mathrm{dB}}^2 \in 
\big[
\bar{\sigma}_{w_\varphi,\mathrm{dB}}^2-\Delta\sigma_{\varphi,\mathrm{dB}},\,
\bar{\sigma}_{w_\varphi,\mathrm{dB}}^2+\Delta\sigma_{\varphi,\mathrm{dB}}
\big], $and $\Delta\sigma_{\varphi,\mathrm{dB}}$ denotes the corresponding uncertainty radius. The corresponding quantities in the linear domain are denoted by $\sigma_{w_\varphi}^2$, $\bar{\sigma}_{w_\varphi}^2$, and $\Delta\sigma_\varphi$, respectively, and they satisfy $
\sigma_{w_\varphi,\mathrm{dB}}^2 = 10\log_{10}\!\left(\sigma_{w_\varphi}^2\right),\quad
\bar{\sigma}_{w_\varphi,\mathrm{dB}}^2 = 10\log_{10}\!\left(\bar{\sigma}_{w_\varphi}^2\right),\quad
\Delta\sigma_{\varphi,\mathrm{dB}} = 10\log_{10}\!\left(\Delta\sigma_\varphi\right),$
where $\Delta\sigma_\varphi>1$ denotes the uncertainty factor in the linear domain.
Accordingly, the probability density function (PDF) of $\sigma_{w_\varphi}^2$ can be expressed as
\begin{equation}
f_{\sigma_{w_\varphi}^2}(x)=
\begin{cases}
\dfrac{1}{2\ln(\Delta\sigma_\varphi)\,x},
&
\dfrac{\bar{\sigma}_{w_\varphi}^2}{\Delta\sigma_\varphi}\le x\le \Delta\sigma_\varphi\bar{\sigma}_{w_\varphi}^2,
\\[2ex]
0, & \text{otherwise}.
\end{cases}
\label{eq:pdf_noise_general}
\end{equation}

\subsection{DEP of $w_r$}

From the perspective of $w_r$, the objective is to determine whether the covert message intended for user $r$ is transmitted. Hence, the received signal at $w_r$ under two hypotheses can be separately expressed as
\begin{equation}
\mathcal{H}_{0}:~
y_{w_r}
=
\mathbf{h}_{w_r}^{H}\mathbf{w}\sqrt{a_t}\,x_t+n_{w_r},
\end{equation}
and
\begin{equation}
\mathcal{H}_{1}:~
y_{w_r}
=
\mathbf{h}_{w_r}^{H}\mathbf{w}
\left(
\sqrt{a_r}\,x_r+\sqrt{a_t}\,x_t
\right)
+n_{w_r},
\end{equation}
where $\mathcal{H}_{0}$ corresponds to the absence of the covert message intended for user $r$, whereas $\mathcal{H}_{1}$ corresponds to its presence.

The DEP of $w_r$ is defined as
\begin{equation}
\zeta_r=P_{{\rm F}}+P_{{\rm M}},
\label{eq:zeta_r_def}
\end{equation}
where $P_{{\rm F}}=\Pr(D_{1}\mid\mathcal{H}_{0})$ and $P_{{\rm M}}=\Pr(D_{0}\mid\mathcal{H}_{1})$ denote the FA and MD probabilities, respectively. Here, $D_{0}$ and $D_{1}$ represent the decisions in favor of hypotheses $\mathcal{H}_{0}$ and $\mathcal{H}_{1}$, respectively. For notational convenience, let $
\Omega_{w_r}\triangleq |\mathbf{h}_{w_r}^{H}\mathbf{w}|^2.$
The FA and MD probabilities can be written as
\begin{equation}
P_{{\rm F}}
=
\Pr(D_{1}\mid\mathcal{H}_{0})
=
\Pr\left\{
a_t\Omega_{w_r}+\sigma_{w_r}^2\ge \Gamma_r
\right\},
\label{eq:PF_def_r}
\end{equation}
and
\begin{equation}
P_{{\rm M}}
=
\Pr(D_{0}\mid\mathcal{H}_{1})
=
\Pr\left\{
\Omega_{w_r}+\sigma_{w_r}^2\le \Gamma_r
\right\},
\label{eq:PM_def_r}
\end{equation}
respectively, where $\Gamma_r$ denotes the detection threshold adopted by $w_r$.

By applying \eqref{eq:pdf_noise_general} to \eqref{eq:PF_def_r} and \eqref{eq:PM_def_r}, the FA and MD probabilities at $w_r$ are respectively obtained as
\begin{equation}
P_{{\rm F}}=
\begin{cases}
1, & \Gamma_r<a_t\Omega_{w_r}+\dfrac{\bar{\sigma}_{w_r}^2}{\Delta\sigma_r},\\[2ex]
\tau_1, & a_t\Omega_{w_r}+\dfrac{\bar{\sigma}_{w_r}^2}{\Delta\sigma_r}\le\Gamma_r<a_t\Omega_{w_r}+\Delta\sigma_r\bar{\sigma}_{w_r}^2,\\[2ex]
0, & \Gamma_r\ge a_t\Omega_{w_r}+\Delta\sigma_r\bar{\sigma}_{w_r}^2,
\end{cases}
\label{eq:PF_r}
\end{equation}

and
\begin{equation}
P_{{\rm M}}=
\begin{cases}
0, & \Gamma_r<\Omega_{w_r}+\dfrac{\bar{\sigma}_{w_r}^2}{\Delta\sigma_r},\\[2ex]
\tau_2, & \Omega_{w_r}+\dfrac{\bar{\sigma}_{w_r}^2}{\Delta\sigma_r}\le\Gamma_r<\Omega_{w_r}+\Delta\sigma_r\bar{\sigma}_{w_r}^2,\\[2ex]
1, & \Gamma_r\ge \Omega_{w_r}+\Delta\sigma_r\bar{\sigma}_{w_r}^2,
\end{cases}
\label{eq:PM_r}
\end{equation}
where $
\tau_1=\frac{1}{2\ln(\Delta\sigma_r)}\ln\!\left(
\frac{\Delta\sigma_r\bar{\sigma}_{w_r}^2}{\Gamma_r-a_t\Omega_{w_r}}
\right)$, $
\tau_2=
\frac{1}{2\ln(\Delta\sigma_r)}
\ln\!\left(
\frac{\Delta\sigma_r(\Gamma_r-\Omega_{w_r})}{\bar{\sigma}_{w_r}^2}
\right).$

\begin{theorem}
\label{thm:DEP_wr}
By substituting \eqref{eq:PF_r} and \eqref{eq:PM_r} into \eqref{eq:zeta_r_def}, the DEP of $w_r$ can be expressed in the following piecewise form. The above piecewise characterization gives rise to two distinct cases because the relative order between the intermediate boundaries $\Omega_{w_r}+\bar{\sigma}_{w_r}^2/\Delta\sigma_r$ and $a_t\Omega_{w_r}+\Delta\sigma_r\bar{\sigma}_{w_r}^2$ will vary with the system parameters.

\textbf{Case 1:} If $
\Omega_{w_r}+\frac{\bar{\sigma}_{w_r}^2}{\Delta\sigma_r}
>
a_t\Omega_{w_r}+\Delta\sigma_r\bar{\sigma}_{w_r}^2,$
then
\begin{equation}
\zeta_r=
\begin{cases}
1, & \Gamma_r<a_t\Omega_{w_r}+\dfrac{\bar{\sigma}_{w_r}^2}{\Delta\sigma_r},\\[2ex]
\tau_1, & a_t\Omega_{w_r}+\dfrac{\bar{\sigma}_{w_r}^2}{\Delta\sigma_r}\le\Gamma_r<a_t\Omega_{w_r}+\Delta\sigma_r\bar{\sigma}_{w_r}^2,\\[2ex]
0, & a_t\Omega_{w_r}+\Delta\sigma_r\bar{\sigma}_{w_r}^2\le\Gamma_r<\Omega_{w_r}+\dfrac{\bar{\sigma}_{w_r}^2}{\Delta\sigma_r},\\[2ex]
\tau_2, & \Omega_{w_r}+\dfrac{\bar{\sigma}_{w_r}^2}{\Delta\sigma_r}\le\Gamma_r<\Omega_{w_r}+\Delta\sigma_r\bar{\sigma}_{w_r}^2,\\[2ex]
1, & \Gamma_r\ge\Omega_{w_r}+\Delta\sigma_r\bar{\sigma}_{w_r}^2.
\end{cases}
\label{eq:zetarcase2}
\end{equation}
\begin{remark}
On the condition of Case 1, there exists a threshold interval for which the DEP becomes zero. In this regime, $w_r$ is able to distinguish between two hypotheses, implying a complete loss of covertness. Accordingly, the subsequent analysis is restricted to Case 2.
\end{remark}

\textbf{Case 2:} If $
\Omega_{w_r}+\frac{\bar{\sigma}_{w_r}^2}{\Delta\sigma_r}
\le
a_t\Omega_{w_r}+\Delta\sigma_r\bar{\sigma}_{w_r}^2,$
then
\begin{equation}
\zeta_r=
\begin{cases}
1, & \Gamma_r<a_t\Omega_{w_r}+\dfrac{\bar{\sigma}_{w_r}^2}{\Delta\sigma_r},\\[2ex]
\tau_1, & a_t\Omega_{w_r}+\dfrac{\bar{\sigma}_{w_r}^2}{\Delta\sigma_r}\le\Gamma_r<\Omega_{w_r}+\dfrac{\bar{\sigma}_{w_r}^2}{\Delta\sigma_r},\\[2ex]
\tau_1+\tau_2, & \Omega_{w_r}+\dfrac{\bar{\sigma}_{w_r}^2}{\Delta\sigma_r}\le\Gamma_r<a_t\Omega_{w_r}+\Delta\sigma_r\bar{\sigma}_{w_r}^2,\\[2ex]
\tau_2, & a_t\Omega_{w_r}+\Delta\sigma_r\bar{\sigma}_{w_r}^2\le\Gamma_r<\Omega_{w_r}+\Delta\sigma_r\bar{\sigma}_{w_r}^2,\\[2ex]
1, & \Gamma_r\ge\Omega_{w_r}+\Delta\sigma_r\bar{\sigma}_{w_r}^2.
\end{cases}
\label{eq:zetarcase1}
\end{equation}
\end{theorem}

\begin{proof}
See Appendix~\ref{app:proof_thm1}.
\end{proof}

\begin{corollary}
\label{cor:DEP_wr_min}
Under Case 2, the optimal detection threshold that minimizes the DEP of $w_r$ is given by
\begin{equation}
\Gamma_r^{opt}=\Omega_{w_r}+\frac{\bar{\sigma}_{w_r}^2}{\Delta\sigma_r}.
\label{eq:Gamma_opt_r}
\end{equation}
The corresponding minimum DEP is
\begin{equation}
\zeta_r^{min}=
\frac{1}{2\ln(\Delta\sigma_r)}
\ln\!\left(
\frac{\Delta\sigma_r^2\bar{\sigma}_{w_r}^2}
{\bar{\sigma}_{w_r}^2+a_r\Delta\sigma_r\Omega_{w_r}}
\right).
\label{eq:zeta_min_r}
\end{equation}
\end{corollary}

\begin{remark}
\label{rem:K_DEP}
Since $K$ denotes the number of PAs, it affects the DEP through the effective leakage term $\Omega_{w_r}$, which determines the decision statistics under $\mathcal{H}_0$ and $\mathcal{H}_1$, named $a_t\Omega_{w_r}+\sigma_{w_r}^2$ and $\Omega_{w_r}+\sigma_{w_r}^2$, respectively. Hence, varying $K$ changes the separability between two hypotheses at $w_r$, and consequently alters not only the DEP level but also the location of the DEP-sensitive power region. The impact of $K$ on covertness should be interpreted in terms of the shift and width of the low DEP interval, rather than solely through a single DEP value.
\end{remark}

\begin{remark}
\label{rem:Delat_DEP}
The DEP does not vary monotonically with $\Delta\sigma_r$. Instead, $\Delta\sigma_r$ affects the DEP primarily by modifying the switching boundaries and the active interval of the piecewise DEP expression, thereby reshaping the DEP profile over the operating power domain. A larger $\Delta\sigma_r$ implies a wider uncertainty set at the warden, which tends to sustain a higher DEP over a broader power region and thus improves the covertness of the considered transmission.
\end{remark}
\subsection{DEP of $w_t$}

Since the DEP analysis for $w_t$ can be conducted in a manner similar to that for $w_r$, only the principal results are provided here. In this case, $x_t$ is regarded as the covert signal, whereas $x_r$ serves as the public signal. Accordingly, the received signal at $w_t$ under two hypotheses can be expressed as
\begin{equation}
\mathcal{H}_{0}^{\star}:~
y_{w_t}
=
\mathbf{h}_{w_t}^{H}\mathbf{w}\sqrt{a_r}\,x_r+n_{w_t},
\label{eq:H0_t}
\end{equation}
and
\begin{equation}
\mathcal{H}_{1}^{\star}:~
y_{w_t}
=
\mathbf{h}_{w_t}^{H}\mathbf{w}
\left(
\sqrt{a_r}\,x_r+\sqrt{a_t}\,x_t
\right)
+n_{w_t},
\label{eq:H1_t}
\end{equation}
where $\mathcal{H}_{0}^{\star}$ and $\mathcal{H}_{1}^{\star}$ correspond to the absence and presence of the covert message intended for user $t$, respectively.

The DEP at $w_t$ is defined as
\begin{equation}
\zeta_t^{\star}=P_{{\rm F}}^{\star}+P_{{\rm M}}^{\star},
\label{eq:zeta_t_star_def}
\end{equation}
where $P_{{\rm F}}^{\star}=\Pr(D_{1}^{\star}\mid\mathcal{H}_{0}^{\star})$ and $P_{{\rm M}}^{\star}=\Pr(D_{0}^{\star}\mid\mathcal{H}_{1}^{\star})$ denote the FA and MD probabilities, respectively. Here, $D_{0}^{\star}$ and $D_{1}^{\star}$ denote the decisions in favor of $\mathcal{H}_{0}^{\star}$ and $\mathcal{H}_{1}^{\star}$, respectively. For notational convenience, define $
\Omega_{w_t}\triangleq |\mathbf{h}_{w_t}^{H}\mathbf{w}|^2.$
Then, the FA and MD probabilities can be written as
\begin{equation}
P_{{\rm F}}^{\star}
=
\Pr\left\{
a_r\Omega_{w_t}+\sigma_{w_t}^2\ge \Gamma_t
\right\},
\label{eq:PF_def_t_star}
\end{equation}
and
\begin{equation}
P_{{\rm M}}^{\star}
=
\Pr\left\{
\Omega_{w_t}+\sigma_{w_t}^2\le \Gamma_t
\right\},
\label{eq:PM_def_t_star}
\end{equation}
respectively, where $\Gamma_t$ denotes the detection threshold adopted by $w_t$.

By substituting \eqref{eq:pdf_noise_general} to \eqref{eq:PF_def_t_star} and \eqref{eq:PM_def_t_star}, the FA and MD probabilities at $w_t$ are respectively obtained as
\begin{equation}
P_{{\rm F}}^{\star}=
\begin{cases}
1, & \Gamma_t<a_r\Omega_{w_t}+\dfrac{\bar{\sigma}_{w_t}^2}{\Delta\sigma_t},\\[2ex]
\tau_3, & a_r\Omega_{w_t}+\dfrac{\bar{\sigma}_{w_t}^2}{\Delta\sigma_t}\le\Gamma_t<a_r\Omega_{w_t}+\Delta\sigma_t\bar{\sigma}_{w_t}^2,\\[2ex]
0, & \Gamma_t\ge a_r\Omega_{w_t}+\Delta\sigma_t\bar{\sigma}_{w_t}^2,
\end{cases}
\label{eq:PF_t_star}
\end{equation}
and
\begin{equation}
P_{{\rm M}}^{\star}=
\begin{cases}
0, & \Gamma_t<\Omega_{w_t}+\dfrac{\bar{\sigma}_{w_t}^2}{\Delta\sigma_t},\\[2ex]
\tau_4, & \Omega_{w_t}+\dfrac{\bar{\sigma}_{w_t}^2}{\Delta\sigma_t}\le\Gamma_t<\Omega_{w_t}+\Delta\sigma_t\bar{\sigma}_{w_t}^2,\\[2ex]
1, & \Gamma_t\ge\Omega_{w_t}+\Delta\sigma_t\bar{\sigma}_{w_t}^2,
\end{cases}
\label{eq:PM_t_star}
\end{equation}
where $
\tau_3=\frac{1}{2\ln(\Delta\sigma_t)}
\ln\!\left(
\frac{\Delta\sigma_t\bar{\sigma}_{w_t}^2}{\Gamma_t-a_r\Omega_{w_t}}
\right)$, $
\tau_4=
\frac{1}{2\ln(\Delta\sigma_t)}
\ln\!\left(
\frac{\Delta\sigma_t(\Gamma_t-\Omega_{w_t})}{\bar{\sigma}_{w_t}^2}
\right).$

\begin{theorem}
By substituting \eqref{eq:PF_t_star} and \eqref{eq:PM_t_star} into \eqref{eq:zeta_t_star_def}, the DEP of $w_t$ is obtained in a piecewise form, determined by the relative ordering of two intermediate detection boundaries.

\textbf{Case 1:} If $
\Omega_{w_t}+\frac{\bar{\sigma}_{w_t}^2}{\Delta\sigma_t}
>
a_r\Omega_{w_t}+\Delta\sigma_t\bar{\sigma}_{w_t}^2,$
then
\begin{equation}
\zeta_t^{\star}=
\begin{cases}
1, & \Gamma_t<a_r\Omega_{w_t}+\dfrac{\bar{\sigma}_{w_t}^2}{\Delta\sigma_t},\\[2ex]
\tau_3, & a_r\Omega_{w_t}+\dfrac{\bar{\sigma}_{w_t}^2}{\Delta\sigma_t}\le\Gamma_t<a_r\Omega_{w_t}+\Delta\sigma_t\bar{\sigma}_{w_t}^2,\\[2ex]
0, & a_r\Omega_{w_t}+\Delta\sigma_t\bar{\sigma}_{w_t}^2\le\Gamma_t<\Omega_{w_t}+\dfrac{\bar{\sigma}_{w_t}^2}{\Delta\sigma_t},\\[2ex]
\tau_4, & \Omega_{w_t}+\dfrac{\bar{\sigma}_{w_t}^2}{\Delta\sigma_t}\le\Gamma_t<\Omega_{w_t}+\Delta\sigma_t\bar{\sigma}_{w_t}^2,\\[2ex]
1, & \Gamma_t\ge\Omega_{w_t}+\Delta\sigma_t\bar{\sigma}_{w_t}^2.
\end{cases}
\label{eq:zeta_t_star_case2}
\end{equation}

\textbf{Case 2:} If $
\Omega_{w_t}+\frac{\bar{\sigma}_{w_t}^2}{\Delta\sigma_t}
\le
a_r\Omega_{w_t}+\Delta\sigma_t\bar{\sigma}_{w_t}^2,$
then
\begin{equation}
\zeta_t^{\star}=
\begin{cases}
1, & \Gamma_t<a_r\Omega_{w_t}+\dfrac{\bar{\sigma}_{w_t}^2}{\Delta\sigma_t},\\[2ex]
\tau_3, & a_r\Omega_{w_t}+\dfrac{\bar{\sigma}_{w_t}^2}{\Delta\sigma_t}\le\Gamma_t<\Omega_{w_t}+\dfrac{\bar{\sigma}_{w_t}^2}{\Delta\sigma_t},\\[2ex]
\tau_3+\tau_4, &\Omega_{w_t}+\dfrac{\bar{\sigma}_{w_t}^2}{\Delta\sigma_t}\le\Gamma_t<a_r\Omega_{w_t}+\Delta\sigma_t\bar{\sigma}_{w_t}^2,\\[2ex]
\tau_4, & a_r\Omega_{w_t}+\Delta\sigma_t\bar{\sigma}_{w_t}^2\le\Gamma_t<\Omega_{w_t}+\Delta\sigma_t\bar{\sigma}_{w_t}^2,\\[2ex]
1, & \Gamma_t\ge\Omega_{w_t}+\Delta\sigma_t\bar{\sigma}_{w_t}^2.
\end{cases}
\label{eq:zeta_t_star_case1}
\end{equation}
\end{theorem}

\begin{proof}
The proof follows analogously to that of the corresponding result for warden $w_r$ and is therefore omitted for brevity.
\end{proof}

\begin{corollary}
\label{cor:DEP_wt_min}
On the condition of Case 2, the optimal detection threshold that minimizes the DEP of $w_t$ is given by
\begin{equation}
\Gamma_t^{\mathrm{opt}}=\Omega_{w_t}+\frac{\bar{\sigma}_{w_t}^2}{\Delta\sigma_t}.
\label{eq:Gamma_opt_t}
\end{equation}
The corresponding minimum DEP is
\begin{equation}
\zeta_t^{\min}=
\frac{1}{2\ln(\Delta\sigma_t)}
\ln\!\left(
\frac{\Delta\sigma_t^2\bar{\sigma}_{w_t}^2}
{\bar{\sigma}_{w_t}^2+a_t\Delta\sigma_t\Omega_{w_t}}
\right).
\label{eq:zeta_min_t}
\end{equation}
\end{corollary}

\section{Covert Transmission Rate}
\label{sec:covert_rate}

The covert transmission rate is defined as the maximum achievable message transmission rate that guarantees reliable communication between the BS and the legitimate users while satisfying the prescribed covertness requirement at the corresponding wardens. In the considered PASS-enabled NOMA covert communication system, user $r$ recovers its own message after SIC, whereas user $t$ decodes its desired message by treating the signal intended for user $r$ as intra-NOMA interference. Due to the location-dependent PASS channel gains, the exact rate expressions are difficult to obtain in closed form. Therefore, numerical quadrature methods are adopted to characterize the covert transmission rates of two users.

\subsection{Covert Transmission Rate of User $r$}

For user $r$ with NISIC, the residual interference caused by non-ideal cancellation should be taken into account. Thus, the covert transmission rate of user $r$ can be expressed as
\begin{equation}
R_{r}^{\rm NISIC}
=
\mathbb{E}
\left[
\log_2
\left(
1+
\frac{a_r\Omega_r}
{\rho |h_I|^2+\sigma_r^2}
\right)
\right],
\label{eq:R_r_NISIC_expectation}
\end{equation}
where $\Omega_r=|\mathbf{h}_r^H\mathbf{w}|^2$ denotes the effective channel gain of user $r$, and $\rho$ is the residual interference power scaling factor.

Due to the location-dependent PASS-NOMA channel gain and the random residual interference, the expectation in \eqref{eq:R_r_NISIC_expectation} does not admit a tractable closed-form expression. Therefore, Gauss--Legendre quadrature is employed for the two-dimensional spatial averaging, whereas Gauss--Laguerre quadrature is applied to the expectation with respect to $|h_I|^2$. Accordingly, the covert transmission rate of user $r$ with NISIC can be approximated as
\begin{equation}
R_{r}^{\rm NISIC}
\approx
\frac{1}{4}
\sum_{i=1}^{N}
\sum_{j=1}^{N}
\sum_{\ell=1}^{L}
\omega_i\omega_j
\omega_{\ell}
\log_2
\left(
1+
\frac{a_r\Omega_{r,ij}}
{\rho\Omega_I u_{\ell}+\sigma_r^2}
\right),
\label{eq:R_r_NISIC_approx}
\end{equation}

where $\{\xi_i,\omega_i\}_{i=1}^{N}$ and $\{\zeta_j,\omega_j\}_{j=1}^{N}$ denote the Gauss--Legendre quadrature nodes and weights associated with the two normalized spatial dimensions, respectively, whereas $\{u_\ell,\omega_\ell\}_{\ell=1}^{L}$ denote the Gauss--Laguerre
quadrature nodes and weights \cite[Secs.~8.5 and 8.6]{Hildebrand1987}. The Gauss--Legendre nodes $\xi_i$ and $\zeta_j$ are mapped to the considered service region through the corresponding affine coordinate transformations, yielding the spatial quadrature point $(x_i,y_j)$. Accordingly, $\Omega_{r,ij}\triangleq\Omega_r(x_i,y_j)$ denotes the effective channel gain of user $r$ evaluated at the $(i,j)$-th spatial quadrature point. Since $|h_I|^2$ is exponentially distributed with mean $\Omega_I$, the substitution $|h_I|^2=\Omega_I u$ converts the residual-interference expectation into the standard Gauss--Laguerre form.

\begin{corollary}
When $\vartheta=0$, the approximated covert transmission rate of user $r$ with ISIC for PASS-NOMA systems is given by
\begin{equation}
R_{r}^{\rm ISIC}
\approx
\frac{1}{4}
\sum_{i=1}^{N}
\sum_{j=1}^{N}
\omega_i\omega_j
\log_2
\left(
1+
\frac{a_r\Omega_{r,ij}}
{\sigma_r^2}
\right).
\label{eq:R_r_ISIC_approx}
\end{equation}
\end{corollary}

\subsection{Covert Transmission Rate of User $t$}

For user $t$, the signal intended for user $r$ is treated as interference. As a result, the covert transmission rate of user $t$ can be expressed as
\begin{equation}
R_t
=
\mathbb{E}
\left[
\log_2
\left(
1+
\frac{a_t\Omega_t}
{a_r\Omega_t+\sigma_t^2}
\right)
\right],
\label{eq:R_t_expectation}
\end{equation}
where $\Omega_t=|\mathbf{h}_t^H\mathbf{w}|^2$ denotes the effective channel gain of user $t$.

Similarly, the closed-form expression of \eqref{eq:R_t_expectation} is generally intractable due to the location-dependent PASS channel gain. By applying Gaussian quadrature to the spatial expectation, the covert transmission rate of user $t$ can be approximated as
\begin{equation}
R_t
\approx
\frac{1}{4}
\sum_{i=1}^{N}
\sum_{j=1}^{N}
\omega_i\omega_j
\log_2
\left(
1+
\frac{a_t\Omega_{t,ij}}
{a_r\Omega_{t,ij}+\sigma_t^2}
\right),
\label{eq:R_t_approx}
\end{equation}
where $\Omega_{t,ij}$ is the effective channel gain of user $t$ evaluated at the spatial quadrature point indexed by $(i,j)$.

\begin{remark}
\label{rem:rate_ceiling}
The derived rate expressions reveal the interference-limited behavior of PASS-NOMA systems. For user $r$, ISIC enables the rate to increase with the effective channel gain, whereas NISIC leads to a throughput ceiling due to residual interference. For user $t$, both the desired signal and the intra-NOMA interference scale with the same effective channel gain, and thus its SINR converges to $a_t/a_r$ in the high-power regime. Accordingly,$
R_t \rightarrow \log_2\left(1+\frac{a_t}{a_r}\right),$
which indicates that the high-power rate of user $t$ is constrained by the NOMA interference structure.
\end{remark}

\section{Numerical Results}
\label{sec:numerical_results}
\suppressfloats[t]

This section presents simulation results to verify the analytical expressions and evaluate the performance of PASS-NOMA. Unless otherwise specified, the simulation parameters are summarized in Table~\ref{tab:simulation_parameters}, following \cite{He2017CL,Xu2025PASSRateMax}. The PAs are uniformly spaced along the waveguide and employ ideal coherent combining to characterize the transmission gain provided by multiple activated PAs. To account for practical SIC imperfections, the residual interference channel gain $|h_I|^2$ under NISIC is modeled as exponentially distributed, with its mean given in Table~\ref{tab:simulation_parameters}.  For the covert rate evaluation, only the transmit-power points satisfying the prescribed covertness requirement at the fixed detection threshold are counted, i.e., the location-averaged DEP of the corresponding warden should be no less than $1-\epsilon$. To facilitate comparison, TA-NOMA and PASS-OMA are adopted as benchmark schemes to evaluate the performance gains achieved by the PASS architecture and NOMA, respectively. For a fair comparison, its transmit power is set equal to the total transmit power of PASS-NOMA, while the other common parameters are kept identical. The transmit power of PASS-OMA is set to half that of PASS-NOMA, i.e., $P_0^{\rm OMA}=0.5P_0$.

\begin{table}[!t]
\centering
\caption{Simulation parameters.}
\label{tab:simulation_parameters}
\renewcommand{\arraystretch}{1.12}
\setlength{\tabcolsep}{5pt}

\renewcommand{\arraystretch}{1.12}
\setlength{\tabcolsep}{3pt}

\begin{tabular}
{|>{\raggedright\arraybackslash}p{0.55\columnwidth}|
 >{\centering\arraybackslash}p{0.32\columnwidth}|}
\hline

\textbf{Parameter} & \textbf{Value} \\ \hline

Power allocation factors for users $r$ and $t$
& \makecell{$a_r=0.3$ \\ $a_t=0.7$} \\ \hline

Number of PAs
& $K=10$ \\ \hline

Carrier frequency
& $f_c=28~\mathrm{GHz}$ \\ \hline

Deployment height
& $d=3~\mathrm{m}$ \\ \hline

Service-region half-widths
& $D_1=D_2=10~\mathrm{m}$ \\ \hline

Path-loss exponent
& $\alpha=2$ \\ \hline

Mean residual interference channel gain
& $\Omega_I=10^{-11}$ \\ \hline

Covertness tolerance
& $\epsilon=0.01$ \\ \hline

\end{tabular}
\end{table}

\begin{figure}[!htbp]
    \centering
    \includegraphics[width=\columnwidth]{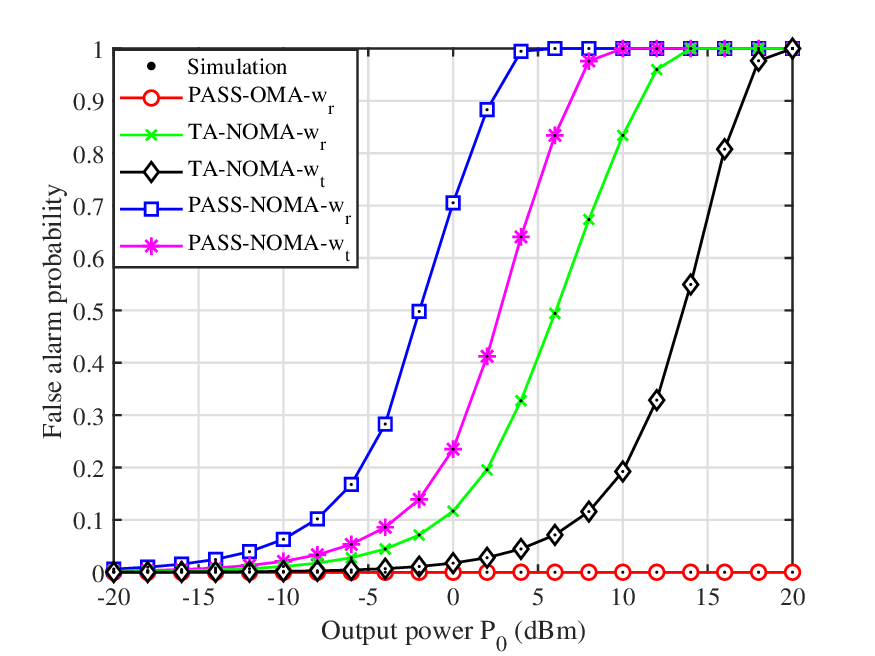}
  \caption{False alarm probability versus output power $P_0$ with $K=10$, $\Gamma_{\varphi}=-68~\mathrm{dBm}$, and $\Delta_{\sigma_{\varphi},\mathrm{dB}}=2~\mathrm{dB}$.}
    \label{fig:PF}
\end{figure}

Fig.~\ref{fig:PF} plots the FA probability versus the output power $P_0$ with $K=10$, $\Gamma_{\varphi}=-68~\mathrm{dBm}$, and $\Delta_{\sigma_{\varphi},\mathrm{dB}}=2~\mathrm{dB}$. The analytical results obtained from \eqref{eq:PF_r} and \eqref{eq:PF_t_star} closely match the Monte Carlo simulation results, validating the derived FA probability expressions. It is evident that the FA probabilities of the NOMA-based schemes increase with $P_0$. The public NOMA signal under $\mathcal{H}_0$ raises the received energy at the wardens and consequently increases the probability of declaring the presence of covert transmission. Moreover, the FA probability at $w_r$ starts to increase at a lower output-power level and rises more rapidly than that at $w_t$. According to \eqref{eq:PF_r} and \eqref{eq:PF_t_star}, this difference originates from the asymmetric public-signal power coefficients, i.e., $a_t>a_r$. In contrast, the FA probability of PASS-OMA remains nearly unchanged with $P_0$ because no public NOMA signal is present under $\mathcal{H}_0$. This contrast highlights the impact of NOMA signal superposition on the FA behavior of the wardens.

\begin{figure}[!htbp]
    \centering
    \includegraphics[width=\columnwidth]{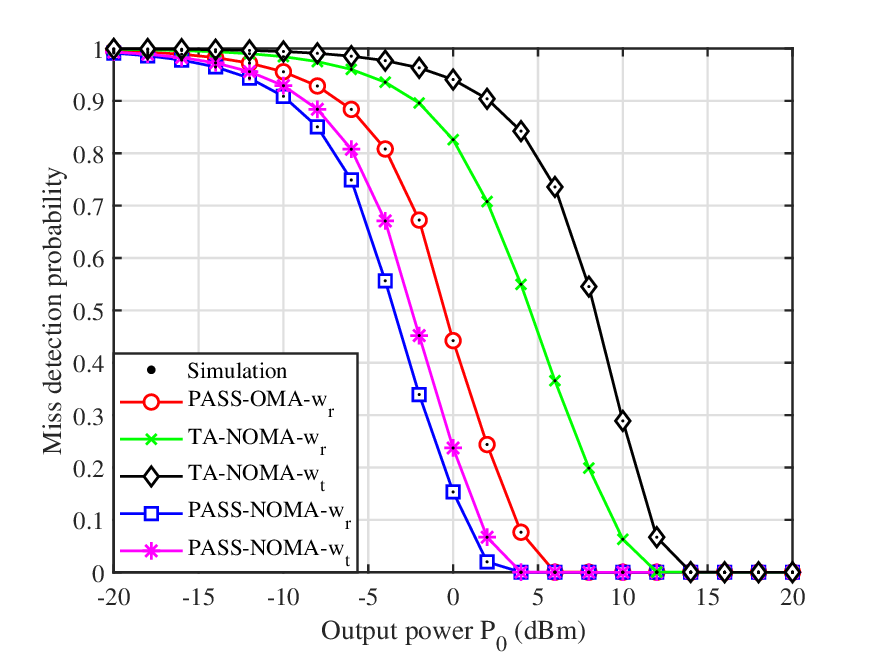}
    \caption{Miss detection probability versus output power $P_0$ with $K=10$, $\Gamma_{\varphi}=-68~\mathrm{dBm}$, and $\Delta_{\sigma_{\varphi},\mathrm{dB}}=2~\mathrm{dB}$.}
    \label{fig:PM}
\end{figure}

Fig.~\ref{fig:PM} illustrates the MD probability versus the output power $P_0$ with $K=10$, $\Gamma_{\varphi}=-68~\mathrm{dBm}$, and $\Delta_{\sigma_{\varphi},\mathrm{dB}}=2~\mathrm{dB}$. The analytical predictions from \eqref{eq:PM_r} and \eqref{eq:PM_t_star} remain in close agreement with the Monte Carlo simulation results. In the low-output-power region, the MD probabilities remain close to one because the small received-energy difference between $\mathcal{H}_0$ and $\mathcal{H}_1$ makes the two hypotheses difficult to distinguish. As $P_0$ increases, the energy separation between the two hypotheses becomes more pronounced, causing the MD probabilities to gradually converge to zero. Compared with TA-NOMA, the MD probabilities of PASS-NOMA begin to decrease at lower output-power levels. Such an earlier transition stems from the coherent superposition of multiple PA radiation paths, which increases the equivalent leakage energy observed by the wardens under $\mathcal{H}_1$.

\begin{figure}[!htbp]
    \centering
    \includegraphics[width=\columnwidth]{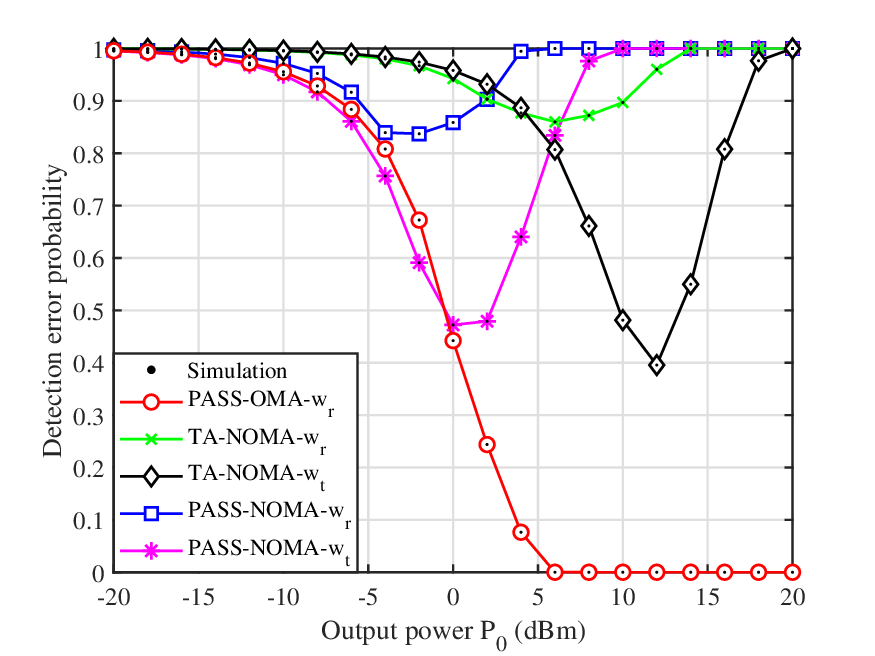}
   \caption{Detection error probability versus output power $P_0$ with $K=10$, $\Gamma_{\varphi}=-68~\mathrm{dBm}$, and $\Delta_{\sigma_{\varphi},\mathrm{dB}}=2~\mathrm{dB}$.}
    \label{fig:DEP_main}
\end{figure}

Fig.~\ref{fig:DEP_main} presents the DEP versus the output power $P_0$ with $K=10$, $\Gamma_{\varphi}=-68~\mathrm{dBm}$, and $\Delta_{\sigma_{\varphi},\mathrm{dB}}=2~\mathrm{dB}$. The Monte Carlo simulation results are in close agreement with the analytical results obtained from \eqref{eq:zetarcase1} and \eqref{eq:zeta_t_star_case1}, confirming the accuracy of the derived DEP expressions. An important observation is that the DEP of the NOMA-based schemes first decreases and then increases with $P_0$. The origin of this valley-shaped profile can be understood from the opposite trends of the FA and MD probabilities shown in Figs.~\ref{fig:PF} and~\ref{fig:PM}. At low output power, the MD probability dominates the DEP. As $P_0$ increases, the MD probability decreases while the FA probability increases, and the latter gradually becomes dominant. This transition is also consistent with the switching behavior of the piecewise DEP expressions in \eqref{eq:zetarcase1} and \eqref{eq:zeta_t_star_case1}. For a fixed detection threshold, varying $P_0$ changes the effective leakage energy and consequently moves the system across different active intervals of the piecewise DEP expressions. PASS-OMA exhibits a different trend, with its DEP decreasing monotonically toward zero because the FA probability remains nearly unchanged in the absence of a public NOMA signal under $\mathcal{H}_0$. Compared with TA-NOMA, PASS-NOMA shifts the DEP valley toward a lower output-power region because coherent PA radiation changes the equivalent leakage energy observed by the wardens. Moreover, under the considered configuration, $w_r$ generally exhibits a higher DEP than $w_t$. This difference is associated with the smaller covert-signal power coefficient $a_r<a_t$, which produces a smaller received-energy increment between the two hypotheses for user $r$, together with the warden-specific leakage gains.

\begin{figure}[!htbp]
    \centering
    \includegraphics[width=\columnwidth]{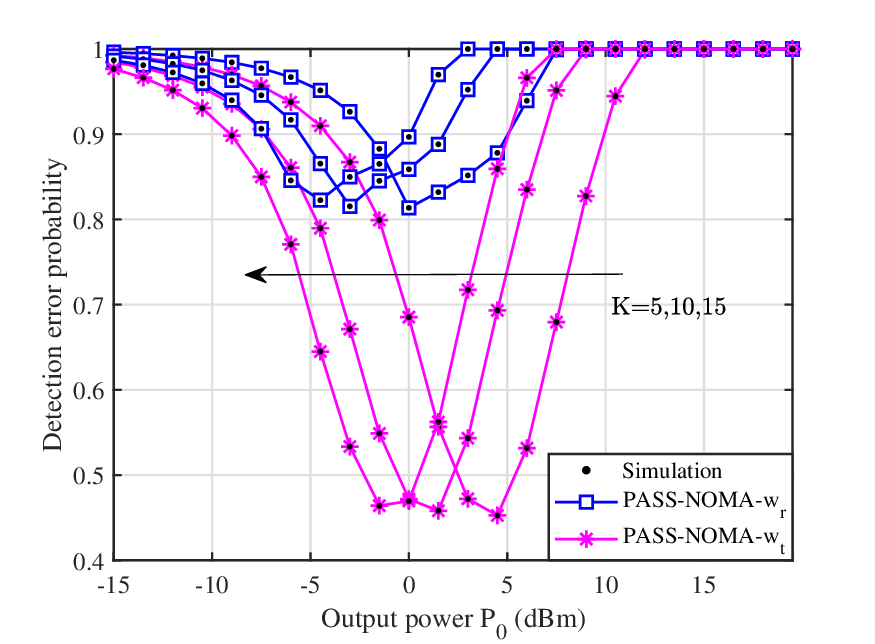}
    \caption{Detection error probability versus output power $P_0$ with $\Gamma_{\varphi}=-68~\mathrm{dBm}$ and $\Delta_{\sigma_{\varphi},\mathrm{dB}}=2~\mathrm{dB}$.}
    \label{fig:DEP_K}
\end{figure}

Fig.~\ref{fig:DEP_K} examines the DEP for different numbers of activated PAs with $\Gamma_{\varphi}=-68~\mathrm{dBm}$ and $\Delta_{\sigma_{\varphi},\mathrm{dB}}=2~\mathrm{dB}$. Increasing $K$ changes not only the DEP level but also the output-power region associated with the DEP valley. In particular, the valleys progressively shift toward lower output-power levels as more PAs are activated. This numerical trend corroborates \Remarkref{rem:K_DEP}, since changing $K$ modifies the effective leakage gain through the coherent superposition of the PA radiation paths. Consequently, the transition regions of the FA and MD probabilities are displaced along the output-power axis. The figure further reveals that the DEP of $w_t$ is more sensitive to the variation of $K$ than that of $w_r$. This difference arises because the larger covert-signal power coefficient $a_t$ makes the energy separation between the two hypotheses at $w_t$ more responsive to variations in the effective leakage gain.

\begin{figure}[!htbp]
    \centering
    \includegraphics[width=\columnwidth]{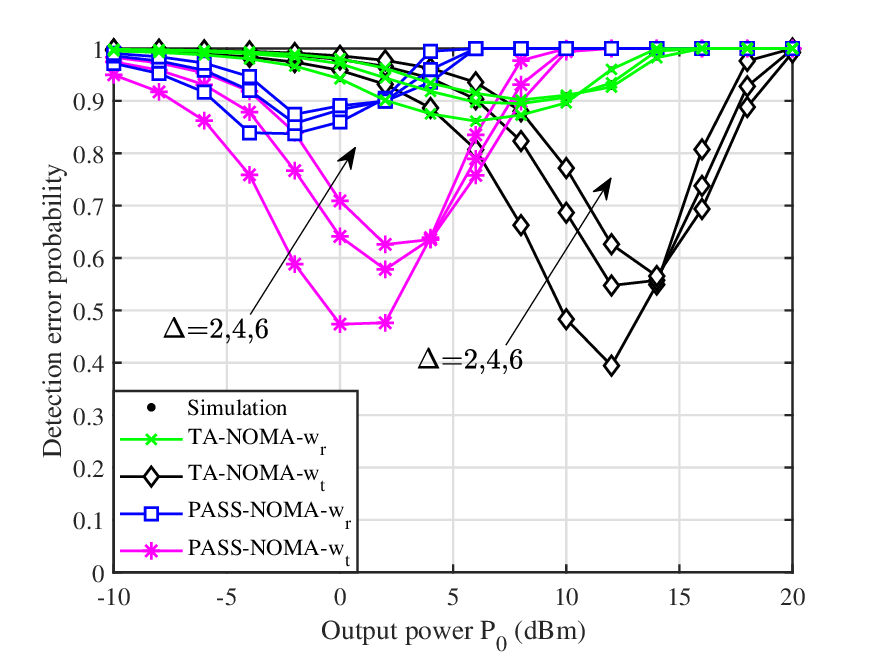}
    \caption{Detection error probability versus output power $P_0$ with $K=10$ and $\Gamma_{\varphi}=-68~\mathrm{dBm}$.}
    \label{fig:DEP_delta}
\end{figure}

Fig.~\ref{fig:DEP_delta} shows the DEP versus the output power $P_0$ for different noise uncertainty factors with $K=10$ and $\Gamma_{\varphi}=-68~\mathrm{dBm}$. Unlike a simple monotonic variation of the DEP at a fixed output power, changing the noise uncertainty reshapes the entire DEP profile. According to \eqref{eq:zetarcase1} and \eqref{eq:zeta_t_star_case1}, enlarging the uncertainty range modifies the switching boundaries and active intervals of the piecewise DEP expressions. Consequently, a larger noise uncertainty tends to sustain higher DEP values over a broader output-power region, which agrees with \Remarkref{rem:Delat_DEP}. Meanwhile, both the location and depth of the DEP valley vary with $\Delta_{\sigma_{\varphi},\mathrm{dB}}$, indicating that the DEP at a particular output power does not necessarily change monotonically with the uncertainty level. In particular, the shallower valleys obtained for larger uncertainty levels reflect a reduced energy-detection capability at the wardens.

\begin{figure}[!htbp]
    \centering
    \includegraphics[width=\columnwidth]{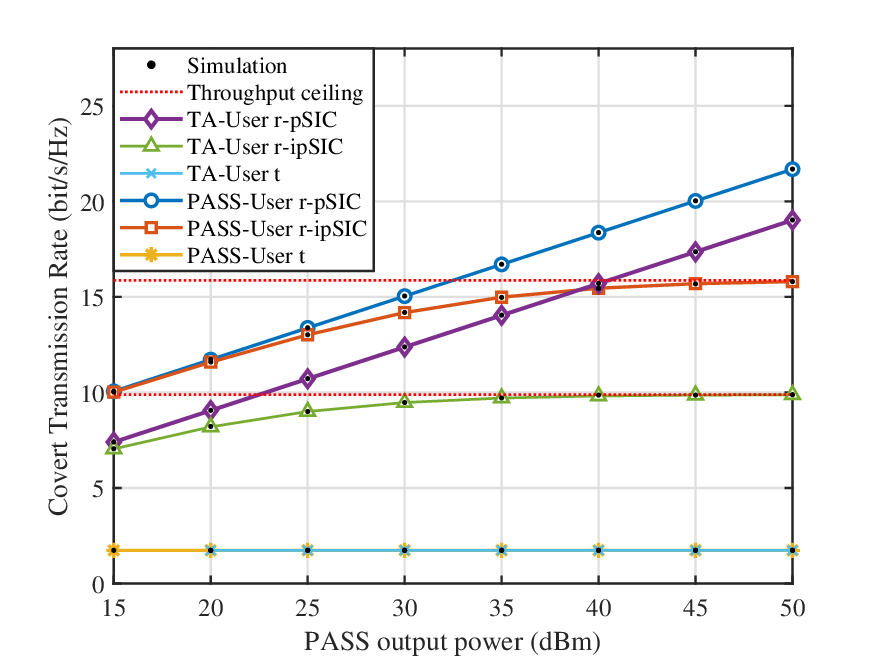}
    \caption{Covert transmission rate versus output power $P_0$ with $K=10$, $\Gamma_{\varphi}=-68~\mathrm{dBm}$, and $\Delta_{\sigma_{\varphi},\mathrm{dB}}=2~\mathrm{dB}$.}
    \label{fig:cssl}
\end{figure}

\Figref{fig:cssl} depicts the covert transmission rate versus the output power $P_0$ with $K=10$, $\Gamma_{\varphi}=-68~\mathrm{dBm}$, and $\Delta_{\sigma_{\varphi},\mathrm{dB}}=2~\mathrm{dB}$ under the prescribed DEP constraint. The analytical approximations in \eqref{eq:R_r_ISIC_approx}, \eqref{eq:R_r_NISIC_approx}, and \eqref{eq:R_t_approx} accurately reproduce the Monte Carlo simulation results. PASS-NOMA provides higher covert transmission rates than TA-NOMA, with the improvement being particularly pronounced for user $r$. The gain mainly arises from the coherent radiation of multiple PAs, which strengthens the effective legitimate channel. For user $r$ with ISIC, removing the inter-user interference allows the covert rate to continue increasing with $P_0$. Under NISIC, however, the residual interference grows with the transmit power and eventually imposes a throughput ceiling. Notably, the covert transmission rate of user $t$ also levels off in the high-power regime because its desired signal and intra-NOMA interference scale with the same effective channel gain. Consequently, its SINR approaches $a_t/a_r$, in accordance with \Remarkref{rem:rate_ceiling}.

\begin{figure}[!htbp]
    \centering
    \includegraphics[width=\columnwidth]{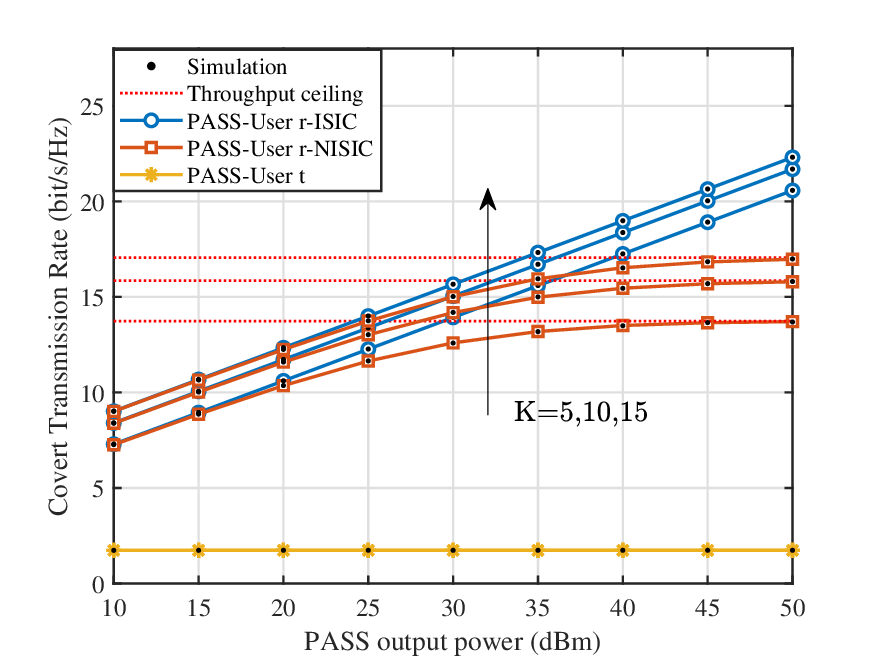}
    \caption{Covert transmission rate versus output power $P_0$ with $\Gamma_{\varphi}=-68~\mathrm{dBm}$ and $\Delta_{\sigma_{\varphi},\mathrm{dB}}=2~\mathrm{dB}$.}
    \label{fig:cssl_k}
\end{figure}

Fig.~\ref{fig:cssl_k} investigates the influence of the number of activated PAs on the covert transmission rate with $\Gamma_{\varphi}=-68~\mathrm{dBm}$ and $\Delta_{\sigma_{\varphi},\mathrm{dB}}=2~\mathrm{dB}$. Increasing $K$ provides a clear rate improvement for user $r$ under both ISIC and NISIC owing to the enhanced effective channel gain. Under ISIC, the separation among the rate curves becomes more pronounced as $P_0$ increases because no residual interference limits the benefit of the additional PA gain. Under NISIC, larger $K$ also raises the achievable rate and its corresponding throughput ceiling, although residual interference still restricts the high-power performance. By comparison, the rate of user $t$ is only weakly affected by $K$ in the medium-to-high output-power region, since the desired signal and intra-NOMA interference are simultaneously strengthened. This result further supports the interference-limited behavior characterized in \Remarkref{rem:rate_ceiling}.

\FloatBarrier

\section{Conclusion}
\label{sec:conclusion}

In this paper, the covert performance of PASS-NOMA communication networks was investigated in the presence of two passive wardens. Under bounded noise power uncertainty, we derived closed-form expressions for the FA and MD probabilities of both wardens and further obtained piecewise closed-form expressions for their DEPs, together with the corresponding optimal detection thresholds and minimum DEPs. The analytical results revealed that the number of activated PAs changes both the DEP level and the location of the DEP valley, while a larger noise uncertainty tends to maintain higher DEP values over a wider output-power region. We further characterized the covert transmission rates of users $r$ and $t$ by considering ISIC and NISIC. Numerical results demonstrated that PASS-NOMA is able to achieve the higher DEP than PASS-OMA and TA-NOMA within specific output-power regions and provides higher covert transmission rates than TA-NOMA under the considered configurations.

\appendices
\renewcommand{\theequation}{\Alph{section}.\arabic{equation}}

\section{Proof of Theorem~\ref{thm:DEP_wr}}
\label{app:proof_thm1}
\setcounter{equation}{0}

From \eqref{eq:PF_r} and \eqref{eq:PM_r}, the DEP of $w_r$ is obtained by summing the FA and MD probabilities, i.e., $\zeta_r=P_{\rm F}+P_{\rm M}$. Since $0<a_t<1$, it follows that $a_t\Omega_{w_r}+\bar{\sigma}_{w_r}^{2}/\Delta\sigma_r<\Omega_{w_r}+\bar{\sigma}_{w_r}^{2}/\Delta\sigma_r$ and $a_t\Omega_{w_r}+\Delta\sigma_r\bar{\sigma}_{w_r}^{2}<\Omega_{w_r}+\Delta\sigma_r\bar{\sigma}_{w_r}^{2}$. Therefore, the final expression of $\zeta_r$ depends on the relative order between $\Omega_{w_r}+\bar{\sigma}_{w_r}^{2}/\Delta\sigma_r$ and $a_t\Omega_{w_r}+\Delta\sigma_r\bar{\sigma}_{w_r}^{2}$.

First, consider the case $\Omega_{w_r}+\bar{\sigma}_{w_r}^{2}/\Delta\sigma_r>a_t\Omega_{w_r}+\Delta\sigma_r\bar{\sigma}_{w_r}^{2}$. In this case, the order of two intermediate boundaries is reversed. Thus, combining \eqref{eq:PF_r} and \eqref{eq:PM_r} yields
\begin{equation}
\zeta_r=
\begin{cases}
1, 
& \Gamma_r<a_t\Omega_{w_r}+\dfrac{\bar{\sigma}_{w_r}^{2}}{\Delta\sigma_r},
\\[2ex]
\tau_1, 
& a_t\Omega_{w_r}+\dfrac{\bar{\sigma}_{w_r}^{2}}{\Delta\sigma_r}
\le \Gamma_r
< a_t\Omega_{w_r}+\Delta\sigma_r\bar{\sigma}_{w_r}^{2},
\\[2ex]
0, 
& a_t\Omega_{w_r}+\Delta\sigma_r\bar{\sigma}_{w_r}^{2}
\le \Gamma_r
< \Omega_{w_r}+\dfrac{\bar{\sigma}_{w_r}^{2}}{\Delta\sigma_r},
\\[2ex]
\tau_2,
& \Omega_{w_r}+\dfrac{\bar{\sigma}_{w_r}^{2}}{\Delta\sigma_r}
\le \Gamma_r
< \Omega_{w_r}+\Delta\sigma_r\bar{\sigma}_{w_r}^{2},
\\[2ex]
1,
& \Gamma_r\ge
\Omega_{w_r}+\Delta\sigma_r\bar{\sigma}_{w_r}^{2}.
\end{cases}
\label{eq:app_case2_zeta}
\end{equation}
In particular, in the interval $a_t\Omega_{w_r}+\Delta\sigma_r\bar{\sigma}_{w_r}^{2}\le \Gamma_r<\Omega_{w_r}+\bar{\sigma}_{w_r}^{2}/\Delta\sigma_r$, both the FA and MD probabilities are zero, i.e., $P_{\rm F}=0$ and $P_{\rm M}=0$, which leads to $\zeta_r=0$. Hence, \eqref{eq:app_case2_zeta} corresponds to the Case 1 expression in \eqref{eq:zetarcase2}.

Next, consider the case $\Omega_{w_r}+\bar{\sigma}_{w_r}^{2}/\Delta\sigma_r\le a_t\Omega_{w_r}+\Delta\sigma_r\bar{\sigma}_{w_r}^{2}$. In this case, the detection threshold $\Gamma_r$ can be divided into five regions. By combining \eqref{eq:PF_r} and \eqref{eq:PM_r}, the DEP is given by
\begin{equation}
\zeta_r=
\begin{cases}
1, 
& \Gamma_r<a_t\Omega_{w_r}+\dfrac{\bar{\sigma}_{w_r}^{2}}{\Delta\sigma_r},
\\[2ex]
\tau_1, 
& a_t\Omega_{w_r}+\dfrac{\bar{\sigma}_{w_r}^{2}}{\Delta\sigma_r}
\le \Gamma_r
< \Omega_{w_r}+\dfrac{\bar{\sigma}_{w_r}^{2}}{\Delta\sigma_r},
\\[2ex]
\tau_1+\tau_2, 
& \Omega_{w_r}+\dfrac{\bar{\sigma}_{w_r}^{2}}{\Delta\sigma_r}
\le \Gamma_r
< a_t\Omega_{w_r}+\Delta\sigma_r\bar{\sigma}_{w_r}^{2},
\\[2ex]
\tau_2,
& a_t\Omega_{w_r}+\Delta\sigma_r\bar{\sigma}_{w_r}^{2}
\le \Gamma_r
< \Omega_{w_r}+\Delta\sigma_r\bar{\sigma}_{w_r}^{2},
\\[2ex]
1,
& \Gamma_r\ge
\Omega_{w_r}+\Delta\sigma_r\bar{\sigma}_{w_r}^{2}.
\end{cases}
\label{eq:app_case1_zeta}
\end{equation}
Hence, \eqref{eq:app_case1_zeta} is equivalent to the Case 2 expression in \eqref{eq:zetarcase1}. This completes the proof.

\section{Proof of Corollary~\ref{cor:DEP_wt_min}}
\label{app:proof_corollary2}
\setcounter{equation}{0}

Under Case 2 of Theorem~2, the DEP of $w_t$ can be regarded as a function of the detection threshold $\Gamma_t$. According to \eqref{eq:zeta_t_star_case1}, the two outer threshold regions yield $\zeta_t^{\star}=1$, which corresponds to the maximum detection error probability and thus cannot minimize the DEP. Therefore, it is sufficient to examine the remaining three nontrivial threshold intervals.

For 
$
a_r\Omega_{w_t}+\bar{\sigma}_{w_t}^{2}/\Delta\sigma_t
\le \Gamma_t
<
\Omega_{w_t}+\bar{\sigma}_{w_t}^{2}/\Delta\sigma_t,
$
the DEP is given by $\zeta_t^{\star}=\tau_3$. Taking the derivative of $\tau_3$ with respect to $\Gamma_t$ yields
\begin{equation}
\frac{d\tau_3}{d\Gamma_t}
=
-\frac{1}
{2\ln(\Delta\sigma_t)\left(\Gamma_t-a_r\Omega_{w_t}\right)}
<0.
\label{eq:app_der_tau3}
\end{equation}
Thus, $\tau_3$ monotonically decreases with $\Gamma_t$ in this interval.

For
$
\Omega_{w_t}+\bar{\sigma}_{w_t}^{2}/\Delta\sigma_t
\le \Gamma_t
<
a_r\Omega_{w_t}+\Delta\sigma_t\bar{\sigma}_{w_t}^{2},
$
the DEP is given by $\zeta_t^{\star}=\tau_3+\tau_4$. Taking the derivative of $\tau_3+\tau_4$ with respect to $\Gamma_t$ gives
\begin{equation}
\frac{d(\tau_3+\tau_4)}{d\Gamma_t}
=
\frac{1}{2\ln(\Delta\sigma_t)}
\left(
\frac{1}{\Gamma_t-\Omega_{w_t}}
-
\frac{1}{\Gamma_t-a_r\Omega_{w_t}}
\right)
>0,
\label{eq:app_der_tau34}
\end{equation}
where the inequality follows from $0<a_r<1$. Hence, $\tau_3+\tau_4$ monotonically increases with $\Gamma_t$ in this interval.

For
$
a_r\Omega_{w_t}+\Delta\sigma_t\bar{\sigma}_{w_t}^{2}
\le \Gamma_t
<
\Omega_{w_t}+\Delta\sigma_t\bar{\sigma}_{w_t}^{2},
$
the DEP is given by $\zeta_t^{\star}=\tau_4$. The derivative of $\tau_4$ with respect to $\Gamma_t$ is
\begin{equation}
\frac{d\tau_4}{d\Gamma_t}
=
\frac{1}
{2\ln(\Delta\sigma_t)\left(\Gamma_t-\Omega_{w_t}\right)}
>0.
\label{eq:app_der_tau4}
\end{equation}
Therefore, $\tau_4$ monotonically increases with $\Gamma_t$ in this interval.

Combining the above monotonicity results, the minimum DEP is achieved at the switching boundary
\begin{equation}
\Gamma_t^{\rm opt}
=
\Omega_{w_t}
+
\frac{\bar{\sigma}_{w_t}^{2}}{\Delta\sigma_t}.
\label{eq:app_Gamma_t_opt}
\end{equation}

Substituting \eqref{eq:app_Gamma_t_opt} into \eqref{eq:zeta_t_star_case1}, the corresponding minimum DEP is obtained as \eqref{eq:zeta_min_t}. 
The proof is completed.
\bibliographystyle{IEEEtran}
\bibliography{mybib}

\end{document}